\documentclass[aps,pra,twocolumn,showpacs,groupedaddress]{revtex4}  
\usepackage{graphicx,color}  
\usepackage{dcolumn}   
\usepackage{bm}        
\usepackage{amssymb}   
\usepackage{subfigure}
\usepackage{sidecap}
\usepackage{comment}

\providecommand{\ignore}[1]{}

\providecommand{\SI}[1]{\,\mathrm{#1}}

\providecommand{\rang}{\rangle}
\providecommand{\lang}{\langle}

\newcommand{\LR}{\textrm{LR}}

\begin{document}

\title{Analysis of Coincidence-Time Loopholes in Experimental Bell Tests}

\author{B. G. Christensen} 
\author{A. Hill}
\author{P. G. Kwiat}
\affiliation{Department of Physics, University of Illinois at Urbana-Champaign, Urbana, IL 61801, USA.}

\author{E. Knill} 
\author{S. W. Nam} 
\author{K. Coakley}
\author{S. Glancy} 
\author{L. K. Shalm} 
\affiliation{National Institute of Standards and Technology, Boulder, CO 80305, USA.}

\author{Y. Zhang} 
\affiliation{Institute for Quantum Computing, University of Waterloo, Waterloo, Ontario N2L 3G1, Canada}

\date{\today}

\begin{abstract}
  We apply a distance-based Bell-test analysis method [E. Knill et
  al., Phys. Rev. A. \textbf{91}, 032105 (2015)] to three experimental
  data sets where conventional analyses failed or required additional
  assumptions.  The first is produced from a new classical source
  exploiting a ``coincidence-time loophole'' for which standard
  analysis falsely shows a Bell violation.  The second is from a
  source previously shown to violate a Bell inequality; the
  distance-based analysis agrees with the previous results but with
  fewer assumptions.  The third data set does not show a violation with
  standard analysis despite the high source quality, but is shown to
  have a strong violation with the distance-based analysis method.
\end{abstract}

\maketitle

Local realism, the notion that any two non-causal events should have
no influence on each other (locality), and that outcomes should be
determined by hidden variables (realism), is fundamental to classical
physics, and is a natural view of reality. When Einstein, Podolsky,
and Rosen noted that quantum mechanics appears to abandon local
realism, they thought that it must be quantum mechanics that is
incomplete~\cite{Einstein1935}. Almost thirty years later, John Bell
showed that local realism and quantum mechanics are not only
conceptually incompatible, but can actually give different statistical
outcomes for experiments on entangled particles~\cite{Bell1964}.  The
statistical differences are quantified via a Bell inequality, a
violation of which would definitively rule out any local realistic
theory, thereby ending a central debate of 20th-century physics.
While entanglement has been experimentally demonstrated in various
physical systems, due to experimental challenges every Bell test to
date has required assumptions about either the source and detector
(e.g., that the detected particles are a fair sample of the total
ensemble emitted from the source), or the possibility of signaling
between specific events (e.g., assuming that there is no signaling
between the measuring devices)\cite{Shadbolt2014}.  While these
assumptions allow one to make arguments against local realism, they
present loopholes that could be exploited by a local realistic model
to violate a Bell inequality.

Furthermore, there can also be implicit assumptions within the data
analysis itself if it directly or tacitly assumes no-signaling or
fair-sampling.  Even worse, the data analysis may directly violate an
assumption, thereby invalidating the analysis technique.  The issue
can be subtle; for example, in the case of the ``coincidence-time
loophole''\cite{Larsson2004}, the implicit assumptions can come from
an otherwise standard coincidence counting method, where the
coincidence windows are centered on one party's detection events (the
implicit assumption is that the local hidden-variable model has no
time-dependence) instead of using a predefined coincidence window.
Finally, additional loopholes can arise from the assumed source
statistics.  Two analysis assumptions are noteworthy.  The first is
that most analyses assume that the source emits particles with
independent and identical states.  The second assumes that the
average violation has a Gaussian distribution; in particular, nearly
all reported Bell violations are cited in terms of numbers of standard
deviations of violation, whose interpretation requires that the
relevant distributions are Gaussian for many standard deviations from
the mean, which fails to hold no matter how many particles are
detected (for a discussion, see~\cite{Zhang2013}).

As Bell tests can be a resource for cryptographic protocols, such as
device-independent random number generation~\cite{Pironio2010} and
device-independent quantum key distribution~\cite{Acin2007}, these
issues are critical to the security of the device, as each loophole
allows for an avenue of attack.  If the device satisfies a
loophole-free Bell test, i.e., violates a Bell inequality with no
extra assumptions, then the device can be trusted regardless of the
manufacturer of the device or possible hacking technique.  Thus it is
important to minimize any extra assumptions required by the analysis
or its interpretation.

In this paper, we begin by describing how common experimental Bell
tests are performed, and the issues that can arise from the data
analysis.  We then briefly summarize a new, distance-based analysis
technique described in Ref.~\cite{KnillTBP}, and in the subsequent
sections we compare this technique to the conventional analysis for real data
sets from three distinct experimental configurations; one is the first demonstration of a system capitalizing on the coincidence-time loophole to fake a Bell-inequality violation, while the other two are a pulsed version and a continuous version of the quantum source presented in Ref.~\cite{Christensen2013}.  Finally, we
discuss general features of the distance-based analysis technique that
apply to all Bell test experiments.

\section{Experimental Bell Tests}
\label{sec:ebt}

An idealized (bipartite) Bell test consists of a series of ``trials''.
For each trial a pair of quantum systems is prepared and distributed
to two parties, Alice and Bob, who independently and randomly choose a
setting at which to measure their quantum system.  For the experiments
discussed here, the quantum systems are photons, and the settings are
chosen from two possibilities, labeled $0$ or $1$.  Alice and Bob's measurement
settings choices are denoted by $s^{A}$ and $s^{B}$, respectively;
their corresponding measurement outcomes are denoted by $t^A_{s^A}$
and $t^B_{s^B}$, respectively.  Since sufficiently high
efficiency detectors are still difficult to obtain, recent experiments
have employed only one detector on each side, so the possible outputs
are either a detection event (which we denote by $t^A_{s^A}=1$ for
Alice, similarly for Bob) or the absence of a detection event (denoted
by $t^A_{s^A}=0$).  After many trials, Alice and Bob compare their
results to determine how the detections are correlated with their
joint settings choices, for example, how often they saw a coincident
detection given a pair of settings.

One type of Bell test experiment in high-efficiency systems uses
polarization entangled photons~\cite{Christensen2013,Giustina2013}.
To generate the entangled photons, a strong pump laser passes through
a nonlinear crystal setup, where each photon of the pump has a
small probability of downconverting into a pair of entangled photons,
one of which is sent to Alice, and the other to Bob.  The measurement
settings for the Bell test are provided by a polarizer placed after either a
half-wave plate in a rotation mount, or a Pockels cell.  Afterwards,
the photons are detected on separate high efficiency photon counters,
e.g., transition-edge-sensor detectors (TES)~\cite{Miller2003}.  The
detection events from each TES are recorded by a time-to-digital
converter, and the resulting timetag sequences are saved for later
analysis to check for correlations.  A new timetag sequence is saved
for each new setting that Alice and Bob choose. Because motorized
rotation mounts (and Pockels cells to a lesser extent) cannot always
change settings quickly, it is possible that multiple detection events
occur before the settings can be changed. For example, the two recent
photon experiments kept the same setting for
$1\SI{s}$~\cite{Christensen2013} and $300\SI{s}$~\cite{Giustina2013}
intervals.  Therefore, multiple conventional one-photon-pair trials
are performed at the same setting, so these trials cannot be
considered strictly independent of each other.  This dependence issue
can be fixed by discarding all but the very first trial for a given
setting (at the cost of much less usable data), or if the data
analysis considers all events taking place while the settings are held
constant to constitute one trial instead of many.  The latter approach
is discussed in the following section.

In the case where a trial is intended to consist of the measurement of
only one photon pair, determining precisely when a given trial occurs
can be difficult.  For example, all single-photon detectors have an
intrinsic uncertainty of the arrival time of the photon.  Furthermore,
downconversion is a probabilistic process, where the emission can
occur at any time when the pump laser has a non-zero amplitude.  This
is most notable for continuous-wave lasers, where downconversion
events happen randomly, uniformly in time.  For each trial Alice and
Bob need to determine if they have a coincidence event (both saw a
detection event), a single event (only one saw a detection), or
neither saw a detection event; they must determine which type of event
occurred despite the temporal uncertainty of their individual events.
(The standard Bell test analysis can be modified so that it is not
necessary to account for the cases where neither party detected a
photon.)  For example, typical quantum optics experiments determine
coincidence events by allowing for a coincidence window around one
party's - say Alice's - detection events: if Bob has a detection event
within the coincidence window determined by her detection event, then
it is called a coincident detection.

In a Bell test, however, this seemingly reasonable method for
determining coincidences cannot exclude all local realistic models, as
it opens up a loophole that can be exploited by a hacker to produce an
apparent Bell inequality violation without any quantum correlations.
The loophole, called the coincidence-time loophole, allows for a
time-dependent local hidden-variable model~\cite{Larsson2004}.
Consider the Clauser-Horne (CH) Bell parameter \cite{Clauser1974} in
the form
\begin{eqnarray}
B_{CH} &=& p_{AB}(t^A_0=1,t^B_0=1)+p_{AB}(t^A_0=1,t^B_1=1)\nonumber
\\ &&{}+p_{AB}(t^A_1=1,t^B_0=1)-p_{AB}(t^A_1=1,t^B_1=1)\nonumber
\\ &&{}-p_{A}(t^A_0=1)-p_{B}(t^B_0=1),
\label{eq:one}
\end{eqnarray}
where $t^{A}_{z}=1$ ($t^{B}_{z}=1$) is a detection event for Alice
(Bob) with $z$ being the measurement setting for Alice's (Bob's)
detector, $p_{AB}(x,y)$ denotes the settings-conditional probability
of the joint outcome of $x$ and $y$ for Alice and Bob's detectors,
respectively, and $p_{A}(x)$ ($p_{B}(x)$) is the setting-conditional
probability of outcome $x$ for any given trial at Alice's (Bob's)
detector. Then it can be shown that $-1\leq B_{CH}\leq 0$ for any
local realistic model.

Consider an experiment where the times of photon-pair arrivals at the
two parties are unknown. Normally, we assume that the pairs arrive at
a constant rate $r_{P}$.  Let $r_{A}(x)$, $r_{B}(x)$ and $r_{AB}(x,y)$
be the rates of events whose detection probabilities are determined by
$p_{A}(x)$, $p_{B}(x)$ and $p_{AB}(x,y)$. Given the constant rate
assumption, we can express $p_{AB}(x,y)= r_{AB}(x,y)/r_{P}$ and
similarly for the other rates. The quantity $B_{CH}$ can be inferred
accordingly and whether or not it is violates the inequality
$B_{CH}\leq 0$ does not depend on the rate $r_{P}$. Thus it is not
necessary to know the rate to observe such a violation.

\begin{figure}
\includegraphics[scale=0.45]{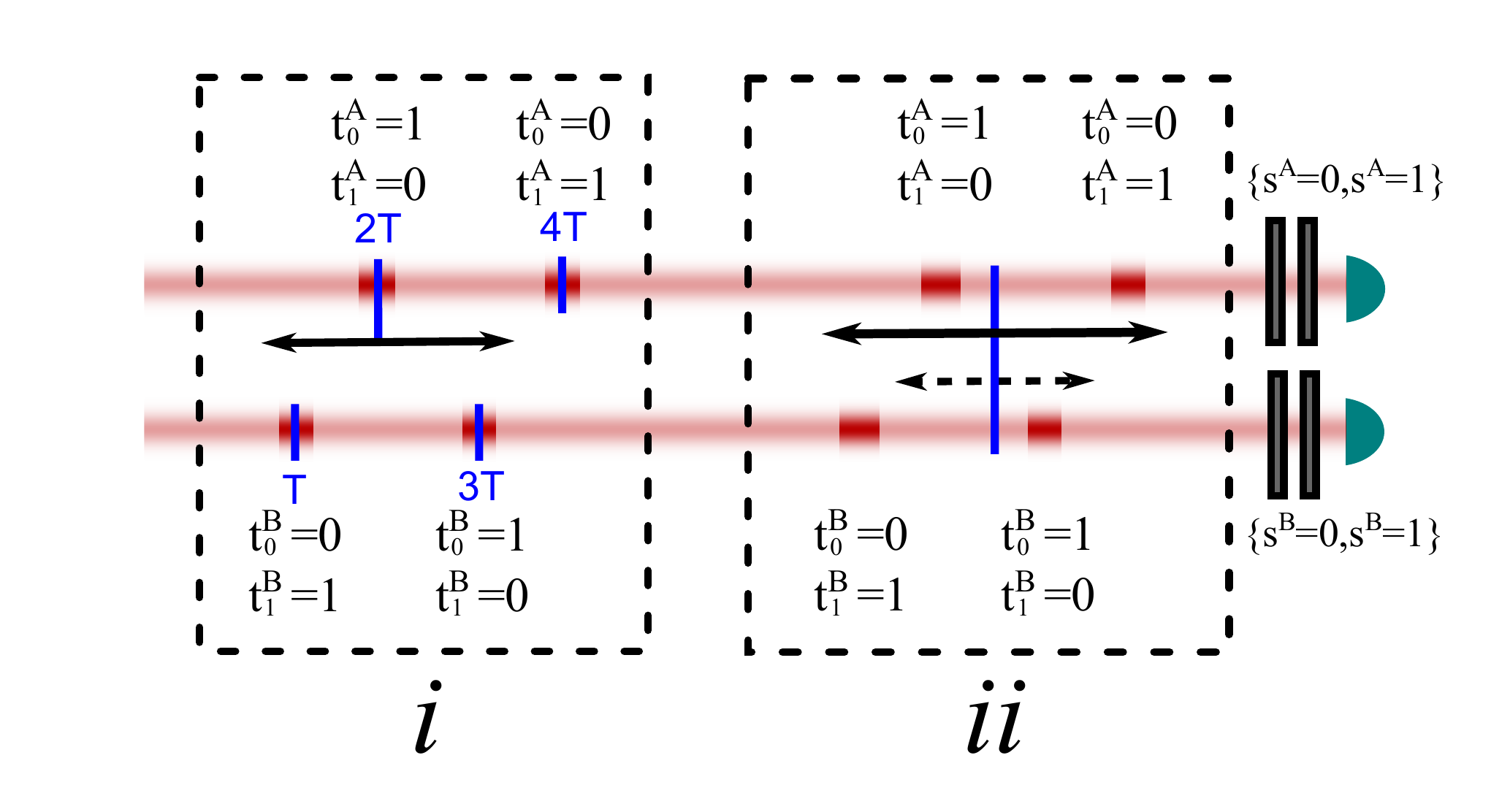}
\caption{\label{fig:cl} A diagram illustrating the coincidence-time
  loophole.  Here $t^A_z = 1$ $(t^A_z = 0)$ corresponds to a detection
  event (no detection event) when Alice chooses measurement setting
  $s^A=z$, and similarly for Bob.  In diagram \textit{i}, a
  coincidence window (black arrow) is selected based on Alice's
  detection event.  A hacker can exploit this loophole by staggering
  pulses in time as shown.  In this case, if the radius (half width)
  of the coincidence window is between $T$ and $3T$, there are no
  $\{s^A=1,s^B=1\}$ coincidence counts, implying that $B_{CH}$ in
  Eq.~\ref{eq:one} is greater than $0$, even for a classical source.
  In diagram \textit{ii}, a well-defined trial is used, where the
  window is centered on a synchronization pulse (blue line); the
  loophole vanishes as there is no longer a way to address only the
  $\{s^A=1,s^B=1\}$ coincidence term.  If a large coincidence window
  (solid arrow) is used, then every measurement setting has a
  coincident event, resulting in $B_{CH}=0$. If a short trial window
  (dashed arrow) is used, then there are only coincident and single events
  at settings $\{s^A=0,s^B=0\}$, giving a Bell value of $B_{CH}=-1$.}
\end{figure}

To exploit the coincidence loophole, a hacker who has full control of
the photon source can, at random times at a rate of $r_{H}$, send a
group of four pulses (two to Alice and two to Bob as shown in
Fig.~\ref{fig:cl}) with each pulse offset by a little less than the
Alice-detection-centered coincidence window used by Alice and Bob.  In
doing so, the pulses that result in detections for settings $s^A=1$
and $s^B=1$ are separated by nearly three coincidence window ``radii''
and therefore do not result in any coincidence counts, whereas at
every other setting combination the detected pulses fall within the
coincidence window.  Consequently, a hacker can achieve an apparent
Bell violation of up to $B_{CH}=r_{H}/r_{P}>0$, given the
experimenter's assumed photon-pair rate $r_{P}$. If the experimenter
attempts to measure $r_{P}$ independently, this measurement may also
be subject to the hacker's manipulations. The trial data set itself only
yields lower bounds on $r_{P}$. That is, assuming (wrongly) that the
detections arise from constant-rate photon pairs, we have that $r_{P}$
should be at least the sum of the rate of detections by $A$ and $B$,
minus the rate of coincident detections at any given setting
combination. This rate is maximized for settings $s^{A}=1$ and
$s^{B}=1$, where it is $2r_{H}$.  Accordingly, $r_{P}\geq
2r_{H}$. Setting $r_{P}=2r_{H}$ gives a maximum inferred violation of
$B_{CH}=1/2$ per presumed pair, which exceeds the maximal quantum
mechanically allowed value of $B_{CH}\approx0.207$ and matches the
maximum allowed by no-signaling~\cite{Popescu1994}.  When quantifying
the violation in Fig.~\ref{fig:cl} and
Fig.~\ref{fig:ctpredefinedwindow} and in the discussion in
App.~\ref{app:clvarsrange}, we use this normalization for the values of
$B_{CH}$, that is we set $r_{P}=2r_{H}$.

One method to prevent the coincidence loophole is to provide
synchronization pulses to both parties, which define the trials
independently of any detection events, as was done in
Ref.~\cite{Christensen2013}.  Another option is to perform a
distance-based analysis of the data, which we now present.

\section{Analysis Description}

A high-level explanation of the coincidence-time loophole is that the
non-local method for inferring trials invalidates the assumptions
underlying the Bell inequalities.  The solution is to ensure that each
party knows in advance the time and duration of a trial and relates
recorded data accordingly. Moreover, if the settings are held fixed
over multiple trials, it is necessary to make additional assumptions;
for example, one can assume that the trials are independent and
identical.  To avoid making such additional assumptions, one should
predefine trials such that they are associated with the time intervals
between making random settings choices.  (An alternative using
party-dependent coincidence window sizes is described in
Ref.~\cite{Larsson2014}.)  For the experiments analyzed here, this
means that each party's measurement outcome is their entire timetag
sequence recorded between making settings choices, rather than a
single detection or non-detection.  Thus, the complete results from a
trial consist of each party's settings choice and the timetag sequences they measured before the next setting was applied.
Note that in the absence of large separations between $A$ and $B$,
this may make it difficult to ensure locality by space-like separation
of relevant events. In two of the experiments below we can exclude all
local realistic probabilistic models, but in principle, a hacker could
have exploited the ability to communicate settings between $A$ and $B$
before the end of a trial to effect arbitrary, non-local-realistic
probability distributions.

Generalizing the notation introduced above, we denote the
timetag-sequence measurement outcomes of the two parties by
$t^{A}_{s^{A}}$ and $t^{B}_{s^{B}}$, where $t^{A}_{s^{A}}$ denotes
$A$'s outcome with the subscript $s^A$ indicating the setting used,
and similarly for $B$. Since the settings choices are under
experimenter control, their probability distribution is known. For the
Bell tests considered here, each of the four setting-choice
combinations has probability $1/4$.

To review the principles of the analysis method in~\cite{KnillTBP},
consider first a general local-realism test.  The method begins by
constructing a Bell function $B$ of trial results such that a Bell
inequality in the form
\begin{equation}
  \label{eq:bellfn_ineq}
  \langle B(t^{A}_{s^{A}},t^{B}_{s^{B}},s^{A},s^{B})\rangle_{\mathrm{LR}} \leq 0
\end{equation}
holds for all local realistic models. Here,
$\langle\ldots\rangle_{\mathrm{LR}}$ denotes the expectation with
respect to a local realistic probability distribution, where the
settings distribution is fixed as above. Given such a Bell function, a
violation can be demonstrated in an experiment by showing a
statistically significant positive value for an empirical estimate
$\hat B$ of $\bar B = \langle
B(t^{A}_{s^{A}},t^{B}_{s^{B}},s_{A},s_{B})\rangle_{\mathrm{EX}}$, where
$\langle\ldots\rangle_{\mathrm{EX}}$ denotes the expectation with
respect to the experimental probability distribution. The traditional
method for evaluating significance is via the sample standard error of
$\hat B$. This can be used to assign approximate confidence intervals
for $\bar B$ but cannot quantify the extremely high significance of
the evidence against local realism that we seek. To quantify the
significance, it is desirable to determine $p$-value bounds in the
framework of statistical hypothesis testing. Ref.~\cite{Zhang2013}
shows how to systematically use lower-bounded Bell functions to obtain
such bounds from the trial results. 

A general strategy for constructing Bell functions that can be used
for conservative estimates of $\bar B$ and $p$-value bounds is given
in Ref.~\cite{KnillTBP}.  Here, ``conservative'' means that the
estimates and bounds are statistically valid with no approximations or
extra assumptions on distributions other than the standard ones,
namely that the settings probabilities are known and that local
realistic distributions are mixtures of outcomes determined by
the local settings. The
fundamental principle is to start with settings-dependent ``distance''
functions $l_{s^A,s^B}(t^A,t^B)$ on the measurement outcome pairs;
such functions are required to satisfy a generalized, twice-iterated
triangle inequality
\begin{equation}
\label{eq:tineq}
  l_{1,1}(t_{1}^{A},t_{1}^{B})
  \leq l_{1,0}(t_{1}^{A},t_{0}^{B})
  + l_{0,0}(t_{0}^{B},t_{0}^{A})
  + l_{0,1}(t_{0}^{A},t_{1}^{B}).
\end{equation}
(If $l$ is non-negative and independent of the settings, then this is the
conventional twice-iterated triangle inequality. Here we use the term
``distance function'' to refer to any function family $l$ satisfying
Eq.~\ref{eq:tineq}.)  Since local realistic models are given by
probability distributions over deterministic models where a party's
setting determines the party's measurement outcome, a Bell function
can be constructed from $l$ according to
\begin{equation}
\label{eq:tbell}
  B(t^{A}_{s^{A}},t^{B}_{s^{B}},s_{A},s_{B}) = 
  \left\{
    \begin{array}{ll}
      l_{1,1}(t^{A}_1,t^{B}_1)& \textrm{if $s_{A}=1$, $s_{B}=1$},\\
    - l_{1,0}(t^{A}_1,t^{B}_0)&\textrm{if $s_{A}=1$, $s_{B}=0$},\\
    - l_{0,0}(t^{B}_0,t^{A}_0)&\textrm{if $s_{A}=0$, $s_{B}=0$},\\
     - l_{0,1}(t^{A}_0,t^{B}_1)&\textrm{if $s_{A}=0$, $s_{B}=1$}.
  \end{array}
  \right.
\end{equation}
The use of distance functions to obtain Bell inequalities was introduced
by Schumacher in Ref.~\cite{schumacher:qc1991a}.

For timetag sequence outcomes associated with experiments that are
intended to violate a CH-type inequality, Ref.~\cite{KnillTBP} shows
that one can define distance functions according to a minimum cost of
converting the first timetag sequence into the second by shifting
and/or deleting timetags. A feature of the technique is that in the
limit where the average time between detections is large compared to
the time-jitter (the uncertainty in the time of the detection), the
distance function can be made to match the value of any CH-type
Bell function. One issue is that the costs defining the distance function
are parametrized, and we wish to choose these parameters optimally
given the characteristics of the experiment. However, to avoid biases
and remain conservative, it is necessary to choose the parameters
beforehand, independent of the data to be analyzed.  That is, contrary
to what is often done in experiments, no part of the ``final data" can
be used to find analysis parameters, such as delays. Otherwise the
validity of confidence intervals or $p$-values is lost. The parameters
can instead be determined by setting aside a fraction of the trials
from the beginning of the experiment.  This ``training data set'' is
used for optimizing analysis parameters.  The remainder of the trials
constitute the analysis data set and should only be analyzed once the
parameters have been chosen. In the applications below, the training
set serves to determine two Bell functions. The first is designed to
maximize a CH-like violation and can be compared to traditional (that
is, non-distance-based) measures of violation.  For reporting these
violations, we modify the conventional method so that the violation
reported is meaningful without assuming that the trials are
independent, as explained in App.~\ref{app:totviol}. The second Bell
function is a systematically ``truncated'' version of the first; the
truncation method is general and can be applied to any distance-based
Bell function~\cite{KnillTBP}.  The second Bell function is bounded,
so we can apply the techniques from Ref.~\cite{Zhang2013} to obtain a
$p$-value (upper) bound. As these $p$-value bounds are extremely
small, we give their negative logarithm base $2$, called the
$\log_2$-$p$-value (lower) bound. See Sect.~\ref{sec:ewv} for the interpretation of
$p$-values and their comparison to Gaussian tails.

Because all the experiments discussed below were performed before the
statistical techniques were fully developed, their analysis was
retrospective and in this sense deviated from the ideal protocol; the
deviations are discussed in App.~\ref{app:discana}.

\section{Experimental realization of the coincidence-time loophole}
\label{sec:excl}

\begin{figure}
\includegraphics[scale=0.5]{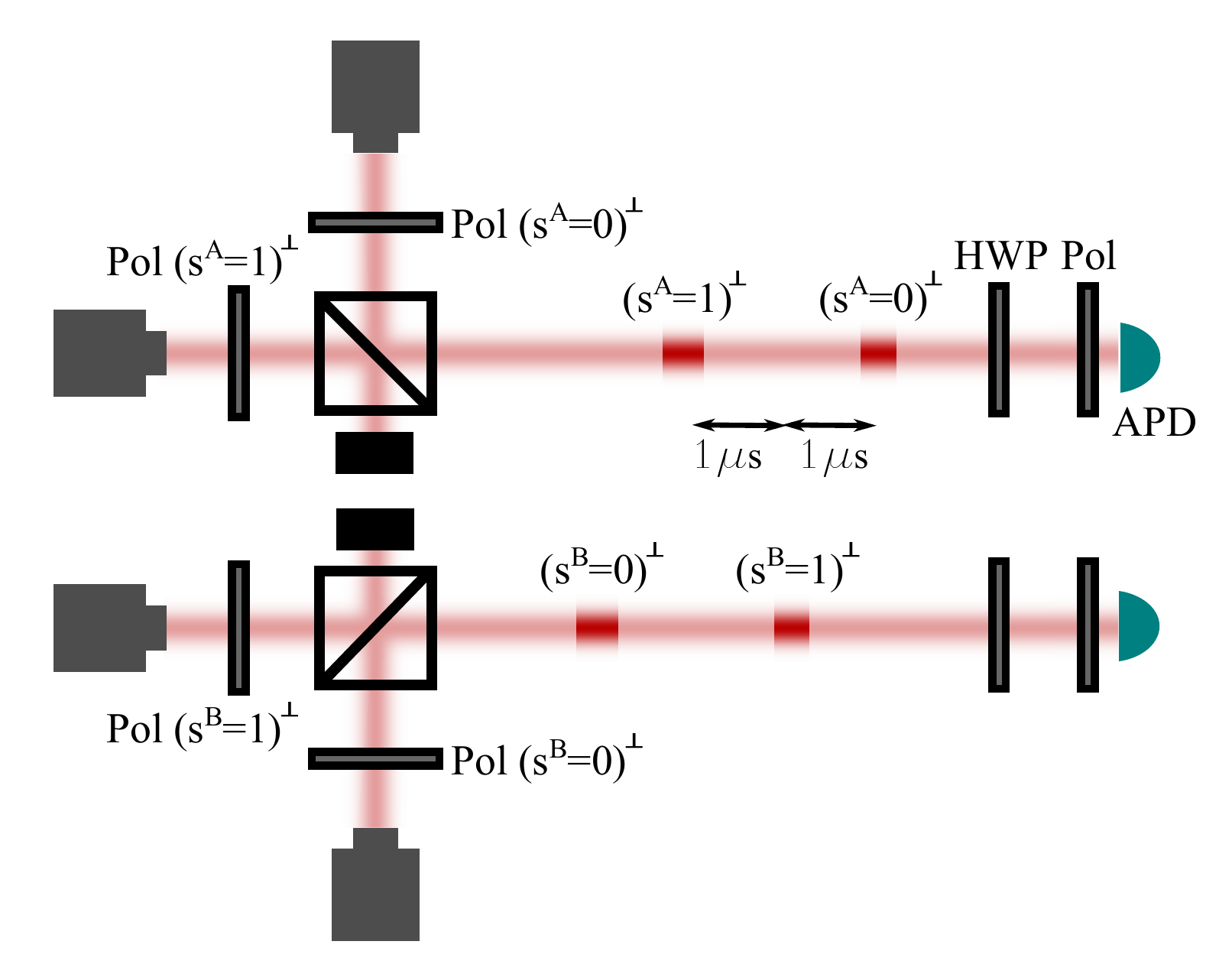}
\caption{\label{fig:second} A diagram of our experimental setup to
  produce the local hidden-variable model described in Fig.~\ref{fig:cl}.
  On Alice's side, we electrically pulse two lasers with a pulse width
  of $100\SI{ns}$; these pulses then pass through polarizers aligned
  orthogonally to her two measurement settings (Pol $(s^A=0)^{\perp}$
  and Pol $(s^A=1)^{\perp}$).  That is, we emit pulses that will
  \textit{not} pass through one of the two measurement settings,
  ensuring only one of the measurement settings will detect our
  optical pulse.  The laser pulse that passes through the
  $(s^A=0)^{\perp}$ polarizer is emitted $2\SI{\mu s}$ before the
  $(s^A=1)^{\perp}$ laser pulse.  We attenuate the lasers enough so
  that after they are combined on a beam splitter, each pulse has a
  mean photon number of approximately $10$, to offset any system loss
  while minimizing the noise due to crosstalk in the polarizers.
  Similarly on Bob's side, we combine two attenuated lasers on a beam
  splitter.  Here, the $(s^B=1)^{\perp}$ pulse is emitted $2\SI{\mu
    s}$ before the $(s^B=0)^{\perp}$ pulse. Both are offset from
  Alice's photon pulses by $1\SI{\mu s}$.  The basis choice for the
  polarization analysis is implemented with a half-wave plate (HWP)
  and polarizer (Pol), where the settings are $-11.25^{o}$ for
  $s^A=0$, $33.75^{o}$ for $s^A=1$, $11.25^{o}$ for $s^B=0$, and
  $-33.75^{o}$ for $s^B=1$ (corresponding to the optimal
  CH-Bell-inequality-violating settings of a perfect maximally
  entangled state).  The photons are then detected by an avalanche
  photo-diode (APD), with an efficiency lower than $66\SI{\%}$, the
  outputs of which are recorded using a time-to-digital converter.
The results of analyzing the data both with a coincidence window
determined by Alice's detection event, as well as a predefined coincidence
window, is displayed in Fig.~\ref{fig:ctpredefinedwindow}.}
\end{figure}

\begin{figure}
\includegraphics[scale=0.38]{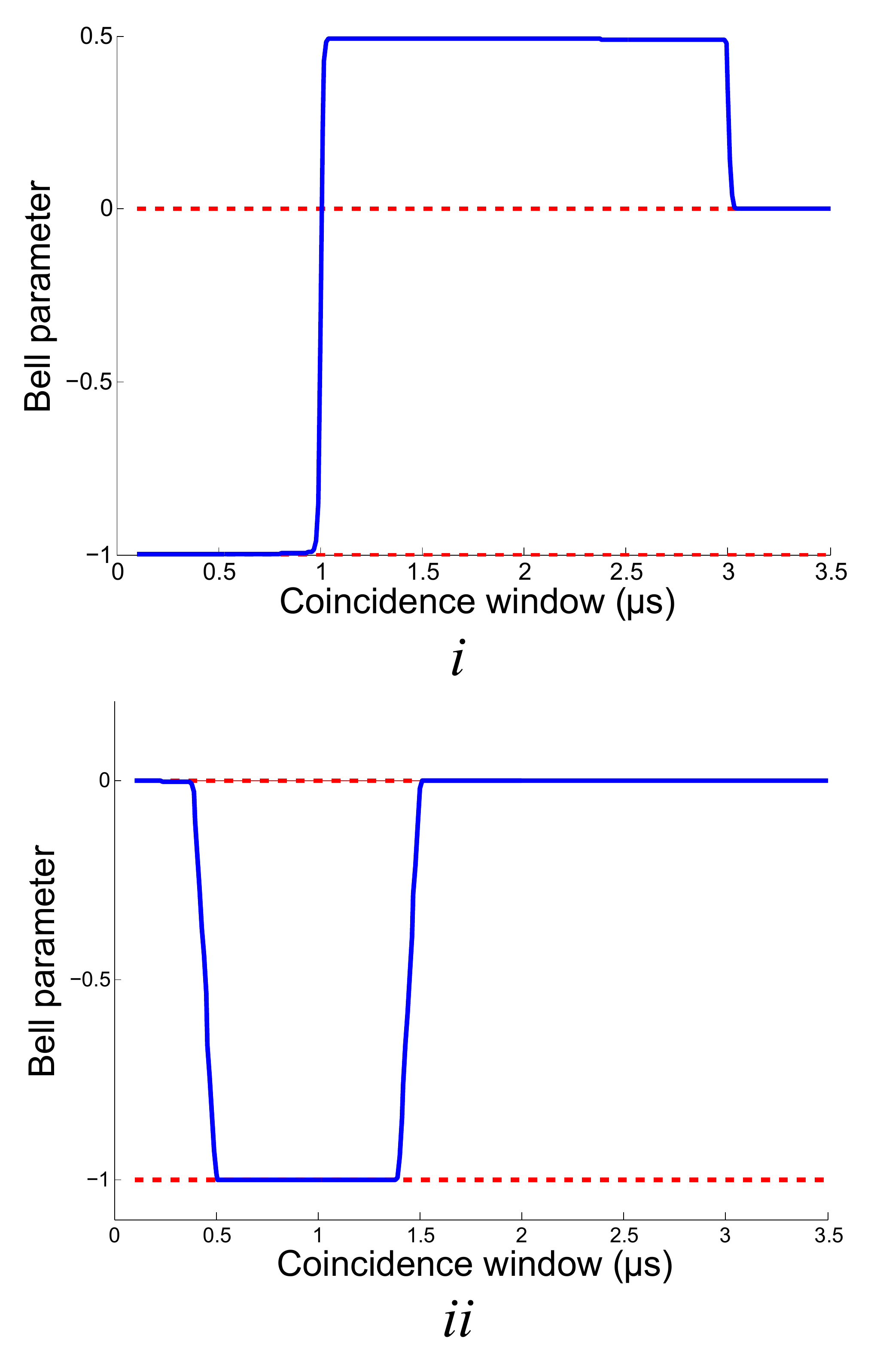}
\caption{\label{fig:ctpredefinedwindow} Two plots of the measured Bell
  parameter, $B_{CH}$ (Eq.~\ref{eq:one}) (solid blue line), as a
  function of the coincidence window radius for our experiment using a
  classical source to produce the local hidden-variable model shown in
  Fig.~\ref{fig:cl}.  When the data set is analyzed with a coincidence
  window determined by a detection event (diagram \textit{i}), 
  the coincidence-time loophole can be exploited to produce a Bell
  violation (values greater than 0).  We separated each pulse by
  $1\SI{\mu s}$, so with this model we see $B_{CH}>0$ for any
  coincidence window radius between $1\SI{\mu s}$ and $3\SI{\mu s}$.
  For coincidence windows less than $1\SI{\mu s}$, we do not have any
  coincidence counts, but we still have single counts, resulting in a
  negative Bell parameter of $B_{CH} \geq -1$.  While this value
  depends on the chosen normalization, the minimal inferred value of
  $r_p$ is at $r_p=2r_h$, resulting in the most negative Bell
  parameter of $-1$.  With window sizes larger than $3\SI{\mu s}$, we
  assign coincident and single events to nearly every detection event,
  resulting in a Bell parameter of $0$.  In contrast, analysis with a
  fixed predefined coincidence window (diagram \textit{ii}), or using
  the technique described in the text, never produces a violation for
  any window size. When the data set is analyzed with a predefined
  coincidence window, the Bell parameter remains between -1 and 0, and
  therefore does not show a violation of local realism.  The results
  match well the predictions given the structure of the classical
  source as explained in App.~\ref{app:clvarspredefined}. The
  positions of the transitions are due to the location of the
  predefined coincidence window relative to the pulse set.  The
  transitions between $0$ and $-1$ are not sharp because of the slow
  desynchronization between the fixed windows and the actual source
  pulse rate.  That is, the window slowly drifts such that it is not
  always centered on the pulse set. For more details, see
  App.~\ref{app:clvarspredefined}.}
 \end{figure}

We realized the coincidence-time loophole experimentally by combining
two attenuated lasers on a beam splitter for both Alice and Bob
(Fig.~\ref{fig:second}).  For Alice, one laser is polarized orthogonal
to the polarizer setting for measurement setting $s^A=0$, while the
other laser is polarized orthogonal to the polarizer setting for
$s^A=1$, and similarly for Bob.  This allows the source to address the
measurement settings independently (i.e., when we send a laser pulse
polarized along $(s^A=1)^{\perp}$, we should only receive detection
events for measurement setting $s^A=0$).  We then attenuate the
sources to a mean photon number per pulse of around $10$. The
relatively high mean photon number offsets the loss in the measurement
and detection process, but is still small enough to minimize the
effect of crosstalk in the polarizer (there is a small chance that the
polarization state to be blocked is still transmitted through the
polarizer).  We then pulse the lasers as shown in Fig.~\ref{fig:cl},
with adjacent pulses separated by $T = 1\SI{\mu s}$.  If we determine
the number of coincidence events by checking if Bob had a detection
event within a window (e.g., $2\SI{\mu s}$) around Alice's detection
events, then we observe Bell inequality violations up to $B_{CH}=0.49$
(with the normalization discussed earlier), where Alice and Bob use
the optimal settings for an ideal maximally entangled state as
specified in the caption of Fig.~\ref{fig:cl}.  A plot of the data
analyzed in this way is displayed in
Fig.~\ref{fig:ctpredefinedwindow}.  We see a ``violation'' of over
$2700$-$\sigma$ (assuming Gaussian statistics).  By altering the two
laser polarizations and increasing the mean photon number to offset
any additional losses, we have been able to exploit this loophole for
a wide range of measurement settings, see App.~\ref{app:clvarsrange}.
In
addition, the degree of violation can be altered by changing the laser
polarization. As a final note, while the plot in
Fig.~\ref{fig:ctpredefinedwindow} has a well-defined structure, it is
possible to broaden the observed ``violation range'' by
probabilistically switching between local hidden-variable models with
different pulse spacings; therefore, one cannot simply look at a plot
of the Bell violation versus coincidence window size to determine if
the coincidence-time loophole is being exploited.

In contrast, if we use a predefined coincidence window centered on a predefined
time rather than one centered on a detection (see App.~\ref{app:clvarspredefined} for details), we do not see a
statistically significant Bell violation, as shown in
Fig.~\ref{fig:ctpredefinedwindow}.
Furthermore, when we use the distance-based analysis from
Ref.~\cite{KnillTBP}, the results correctly do not indicate that the
system is behaving contrary to local realism.  Initially, in the
training set, where delays are determined to offset electronic
latencies, the delays on apparent coincidences were found to depend
highly on the measurement settings, due to the scheme for
exploiting the coincidence loophole.  From the other experiments using
the same setup, the latencies are known to be small, so for
demonstration purposes we ignored this observation and did not offset
for electronic latencies.  The distance-based Bell function
(Eq.~\ref{eq:tbell}) is then significantly negative, showing no
evidence against local realism according to this analysis.

While the data set in this case is contrived to be clearly determined by a local
hidden-variable model, in real experiments the issues are far more
subtle.  For example, avalanche photo-diodes can have a
count-rate-dependent latency, and since each measurement setting can
have different detection rates (for example, in
Ref.~\cite{Christensen2013}, the count rates differed by a factor of
3), it is critical that the analysis is not susceptible to these minor
latency shifts.  To show that these issues are relevant,
Ref.~\cite{KnillTBP} presents a coincidence-loophole-exploiting scheme
whose statistics closely match those of a standard photon-pair source.

\section{Experiments with violation}
\label{sec:ewv}

The example above  shows the use of the distance-based analysis
technique to ``catch'' an invalid violation of a Bell inequality with
a purely classical source.  The following two examples demonstrate the
strength of this analysis on data with actual quantum correlations.

First, we consider the data collected and analyzed in
Ref.~\cite{Christensen2013}, where the experiment had a high enough
system efficiency and low enough noise to be able to violate a CH Bell
inequality.  The data set was taken with an external clock
synchronized to a laser pulse. A predefined coincidence window of
$2.4\SI{\mu s}$ was centered around the laser synchronization pulse,
from which a trial could be well defined, avoiding the
coincidence-time loophole discussed above.  The data set was collected
by changing the measurement settings randomly every second, collecting
for $4450$ different measurement setting choices.  For the analysis in
Ref.~\cite{Christensen2013}, the data set was partitioned into $50$
different Bell tests.  The uncertainty was calculated from the
distribution of the $50$ different Bell parameters using the sample
standard error.  The reported value from this approach was
$B_{CH}=5.4\times 10^{-5}\pm 7.0\times 10^{-6}$, a $7.7$-$\sigma$
violation, where the conventional interpretation of the large
violation assumes Gaussian statistics.  In contrast, here we analyze
the same data set with the distance-based method without making
distributional assumptions or approximations.  Additionally, whereas
the previous analysis required each pulsed trial to be independent and
identical to allow for the settings being fixed, the new analysis,
detailed in App.~\ref{app:discana}, treats each $1\SI{s}$ period with
fixed settings as one trial and therefore does not require this
assumption.  We find a $\log_{2}$-$p$-value bound of $33$, which means
that for every local realistic model, the probability that this
analysis reports a $\log_{2}$-$p$-value above $33$ is less than
$2^{-33}=1.16\times 10^{-10}$, a very unlikely event.  While this
result is equivalent to a $6.3$-$\sigma$ violation for Gaussian
statistics (we give the Gaussian-equivalent violation only for
comparison; it is computed from the $p$-value bound of $2^{-33}$ by
solving $\int_{x}^{\infty}e^{-x^{2}/2}/\sqrt{2\pi}=2^{-33}$ for $x$),
slightly lower than the $7.7$-$\sigma$ violation reported in Ref. \cite{Christensen2013},
it does not assume Gaussian statistics.  Thus, we see that with
minimal degradation of the evidence for Bell-inequality violation, we
have reduced the required assumptions on the system: the trials do not
need to be independent and identical and the distributions are not
approximated by Gaussians. If the system is being hacked, lack of
independence and Gaussianity are even more pronounced.

\begin{figure}
\includegraphics[scale=0.38]{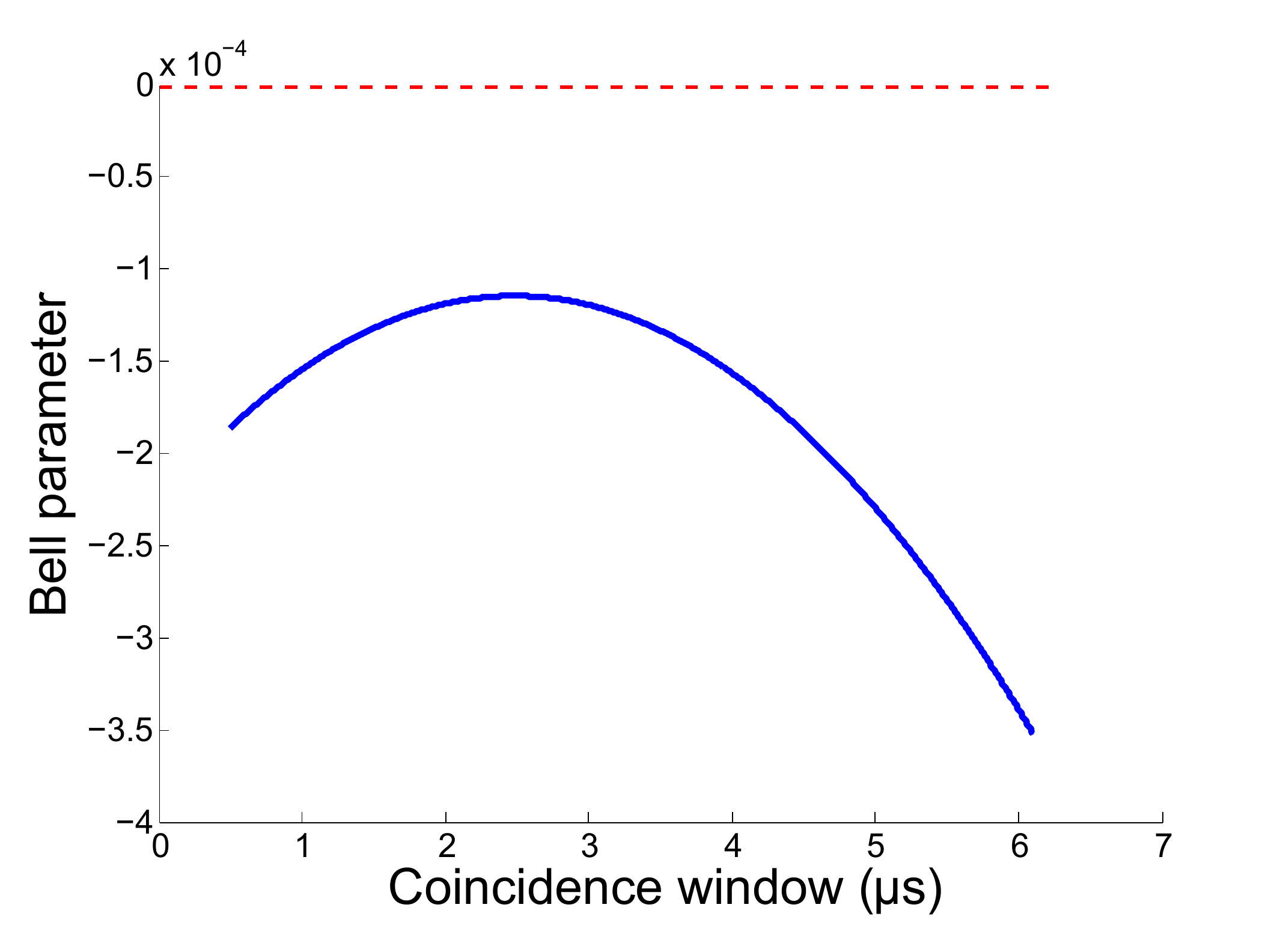}
\caption{\label{fig:fourth} A plot of the CH-Bell parameter from a
  non-pulsed experiment, analyzed using different predefined coincidence
  windows.  While the system is capable of a detection loophole-free
  violation as verified in Ref. \cite{Christensen2013}, the
  inefficiencies of a conventional analysis with predefined
  coincidence windows are unable to produce a non-classical result:
  for small window sizes, the analysis introduces loss because of the
  timing uncertainty (time-jitter) of the detectors, causing the
  mis-assignment of some detection events as non-coincidence counts;
  large windows increase the system noise, to which the CH Bell
  inequality is very sensitive, again resulting in a reduced Bell
  parameter.  Here we show the Bell parameter for many different
  coincidence windows, the blue solid line is a fit to all of the data
  points (the points are spaced by $10\SI{ns}$), each of which lie within the
  thickness of the line.  With conventional analysis, we do not
  observe a Bell violation (above the red dashed line) for any
  coincidence window.  With the new analysis discussed in
  \cite{KnillTBP}, we observe a violation with a $\log_2$-$p$-value
  bound of $269$.}
\end{figure}

Finally, we consider a different data set taken on the same
high-efficiency system, but with neither pulsing the laser nor having an
external clock.  To analyze the data conventionally, we partition time
into segments independent of the data (that is, we impose a fixed
coincidence window).  Since we are not determining the coincidence
window based on the data, it is not susceptible to the
coincidence-time loophole (see Fig.~\ref{fig:cl}\textit{ii}).
However, because we are introducing a coincidence window that is not
related to the arrival time of photons, due to the detector
time-jitter, we are effectively introducing loss into the system.  That
is, if the window is small compared to the time-jitter, then it is
possible that Alice and Bob's detection events from a single pair of
photons are nevertheless registered in different time segments,
resulting in two single counts without a coincidence count.  In the
opposite limit, the window becomes too large, which reduces the Bell
parameter due to the high likelihood of counting uncorrelated photon
pairs as coincidences.  The result of analyzing the data in this way
is displayed in Fig.~\ref{fig:fourth}.  While the source quality is
sufficient for a Bell test (that is, it has high heralding efficiency
and high entanglement quality), the effective loss introduced by this
conventional analysis is too much for us to adequately extract the
quantum correlations. If we instead analyze the data using the
distance-based approach discussed here, we find a violation with a
$\log_2$-$p$-value bound of $269$, the equivalent of a $19$-$\sigma$
violation (see App.~\ref{app:discana} for more details). In addition
to revealing a violation where conventional analysis would not produce
one, the confidence in the violation is actually significantly larger
than that with the pulsed source presented in
Ref.~\cite{Christensen2013}.  This is because we can utilize a system
that is ``on'' more often than a pulsed source (which by definition
has no data collection between pulses), thereby resulting in
substantially more data.

\section{Discussion}

As shown in the above examples, the distance-based analysis of
Ref. \cite{KnillTBP} is able to improve the statistical significance
of a Bell inequality violation, as well as reduce the required
assumptions compared to a standard analysis.  While the analysis uses
distance functions as a measure of the violation, it has important
features common to any conservative analysis of Bell inequality data.
First, to estimate the significance of the violation, it is important
to use $p$-value bounds instead of standard deviations. The latter are
unreliable for the high significance of typical Bell inequality
violations.  Second, to prevent overestimating the statistical
significance of the Bell inequality violation, delays, coincidence
windows and other such analysis parameters should be determined from a
training data set (that is then discarded) rather than the data to be
analyzed.  Otherwise, if the final data set is used to determine these
parameters, the reported violation may be biased by statistical
fluctuation rather than reflect a fair estimate.  Finally, all Bell
tests should have predefined trials to avoid opening up additional
loopholes (e.g., the coincidence-time loophole). The predefined trials
may be based on a timetag sequence according to the chosen settings as
presented here, specific laser pulses detected on a photodiode as
presented in Ref.~\cite{Christensen2013}, or the detection of
heralding photons as in the ion experiments of
Ref.~\cite{Pironio2010}.

\begin{acknowledgments}
  The authors thank K. T. McCusker, J. B. Altepeter, B. Calkins,
  T. Gerrits, A. E. Lita, and A. Miller for assistance with the
  quantum source.  This work includes contributions of the National
  Institute of Standards and Technology, which are not subject to
  U.S. copyright.  This research was supported by the NSF grant
  No. PHY 12-05870.
\end{acknowledgments}

\appendix

\section{Additional Experimental Details}
\label{app:clvars}

This section further discusses our classical source that exploits the
coincidence time loophole.  The first subsection explains how the
source can be tuned to match Alice and Bob's expectations (i.e., to
give violations consistent with quantum mechanics).  In the second
subsection we use a predefined coincidence window to analyze the data
from the classical source and find no violation of local realism.

\subsection{Controlling Violation Size}
\label{app:clvarsrange}

\ignore{The Tsirelson bound for the CHSH inequality is
\begin{equation}
\langle A_{0}B_{1}\rangle +
\langle A_{0}B_{0}\rangle +
\langle A_{1}B_{0}\rangle -
\langle A_{1}B_{1}\rangle \leq 2\sqrt{2}.
\end{equation}
Here the variables are $\{-1,1\}$-valued.  
Define $a=(A+1)/2$, $b=(B+1)/2$) and for $d(a,b)=|a-b|$, use
$\langle A B\rangle = \langle 1-2d(a,b)\rangle$ to get
\begin{equation}
\langle d(a_{1},b_{1})\rangle
-\langle d(a_{0},b_{1})\rangle
-\langle d(a_{0},b_{0})\rangle
-\langle d(a_{1},b_{0})\rangle
\leq \sqrt{2}-1.
\end{equation}
The left-hand side here is the sum of a CH Bell parameter and its A/B
reflection. Up to no signaling, it is therefore twice the CH Bell
parameter with respect to nominal trials, or identical to the CH Bell
parameter with respect to the hacker's trials.}

\begin{figure}
\includegraphics[scale=0.38]{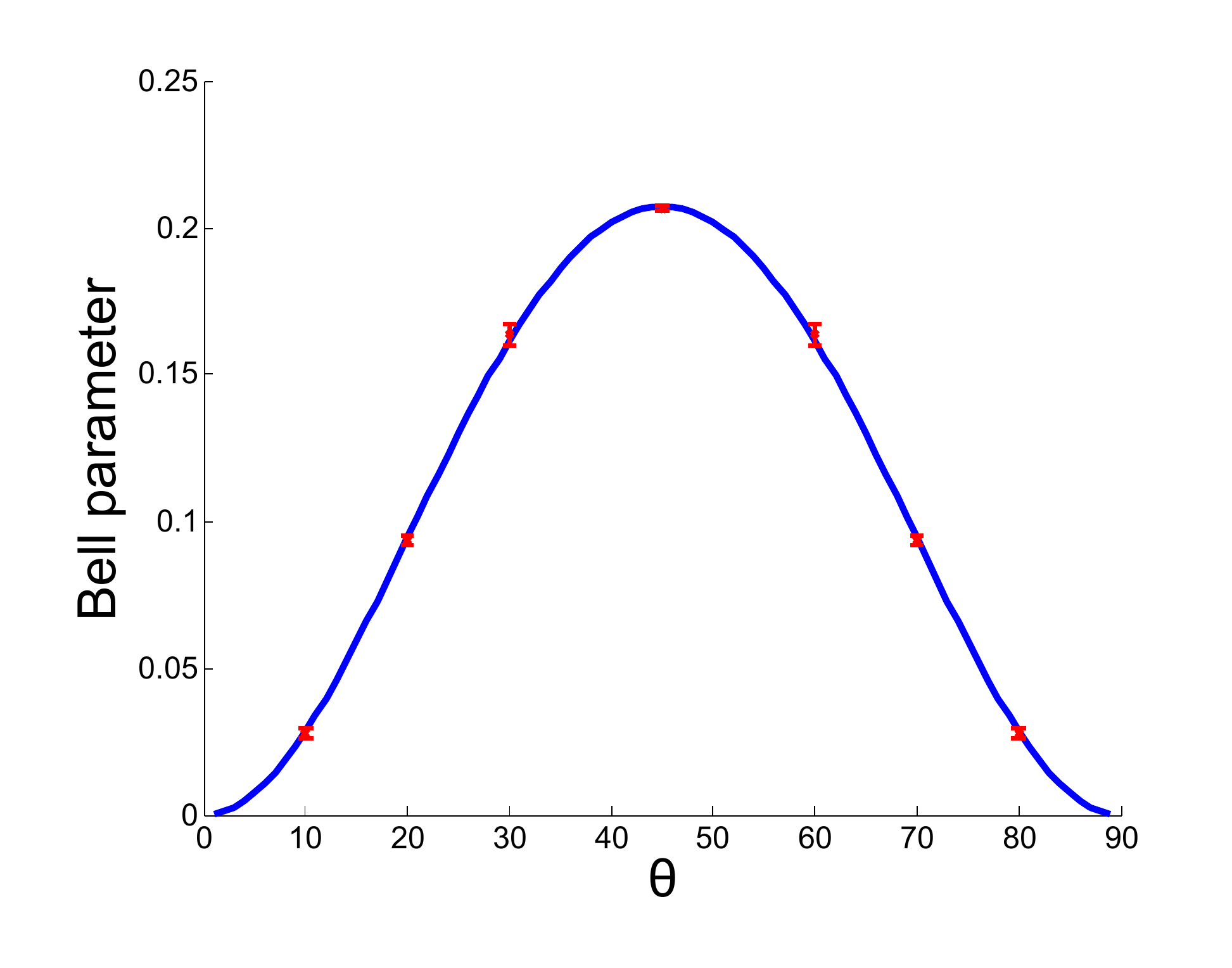}
\caption{\label{fig:fakedatatuning} A plot of a few measured data
  points from our classical source when analyzed incorrectly (susceptible to the coincidence-time loophole discussed in Sect.~\ref{sec:ebt}).  The
  blue curve is the predicted quantum mechanical maximum given the
  state $\cos{\theta}|HH \rangle + \sin{\theta} |VV \rangle$.  Here,
  we assumed Alice and Bob have a target $\theta$ and use the optimal
  measurement settings for the input state.  We then adjusted the
  input polarizer angle to match the quantum mechanically allowed
  maximum Bell parameter, given Alice and Bob's measurement settings.
  The results of a few measurements are displayed as the red data
  points, which are indistinguishable from the quantum mechanical expectation.}
\end{figure}

In an actual attempt of a Bell test, Alice and Bob would likely
suspect the presence of a hacker if their estimated CH-Bell parameter
is beyond the quantum mechanical limit of $(\sqrt{2}-1)/2$.  
Even more so, if Alice and Bob know that they have low system
efficiencies, then the value they expect is well below $(\sqrt{2}-1)/2$.  In
particular, with low efficiency, Alice and Bob design their system to
use states of the form $\cos{\theta}|HH \rangle + \sin{\theta} |VV
\rangle$ (see
Ref.~\cite{eberhard:qc1993a,Giustina2013,Christensen2013}) to maximize
the measured violation.  Consequently, a hacker would want Alice and Bob to believe that they prepared a less
entangled state (states with $\theta$ farther from $\pi/4$).  If Alice
and Bob estimate $\theta$ for their state, they obtain a maximum Bell
parameter they expect.  Ideally, the hacker controls the measured Bell
parameter to match Alice and Bob's expectation and avoid suspicion.
In our case, with the source depicted in Fig.~\ref{fig:second}, we can
tune the input polarizers (and adjust the laser diode brightness to
compensate the increased loss) to create nearly any value of the Bell
parameter.  The results of a few measurements using this technique are
displayed in Fig.~\ref{fig:fakedatatuning}.

\subsection{Predefined Window Analysis}
\label{app:clvarspredefined}

To use a predefined window to analyze the data exploiting the
coincidence-time loophole, we first add in a synchronization signal at
the rate equal to the rate that the source emits a set of pulses, 100
kHz in our case.  As there was no actual synchronization signal when
the data set was taken, we implement this signal in post-processing.  For
comparison with Fig.~\ref{fig:cl}\textit{ii}, where the predefined
coincidence window is in the center of the pulse set, we placed the
first synchronization signal in the center as determined by the first
two detection events in the data set.  We then create a periodic
signal by spacing each synchronization signal by 10 $\mu$s ($=1/100$
kHz).  To compensate for the relative temporal drift between the
function generator and the timetagging electronics, we reset the
synchronization signal every 500 detection events to be re-centered in
the pulse set.  If the separation between adjacent pulses is 1 $\mu$s
(see Fig.~\ref{fig:second}), then we would expect a Bell parameter
close to 0 for windows less than 0.5 $\mu$s, since there will be
neither single nor coincident events (other than occasional dark
counts, no event will fall within the predefined window).  For windows
between $0.5\SI{\mu s}$ and $1.5\SI{\mu s}$ we would expect a Bell parameter
close to $-1$, since we see events primarily from $\{s^A=0,s^B=0\}$.
That is, $p_{AB}(t^A_0=1,t^B_0=1)=1$, $p_{A}(t^A_0=1)=1$, and
$p_{B}(t^B_0=1)=1$ in Eq.~\ref{eq:one}, while all other terms are 0.
Finally, for predefined window sizes larger than 1.5 $\mu$s, all terms
in Eq.~\ref{eq:one} are equal to $1$, leading to a Bell parameter of 0.
The results of analyzing the classical data with a predefined
coincidence window of variable width are displayed in
Fig.~\ref{fig:ctpredefinedwindow}.

\section{Discussion of Analyses}
\label{app:discana}
  
Here we describe in detail the distance-based analyses of the data
from the three experiments discussed in the paper. The results
reported are from final analyses that adhered to the protocol of
inferring parameters from the training set and applying them
adaptively to the analysis set. However, the final analyses were not
strictly blind; the data set was available for some time while our
analysis methods were being developed and there were multiple early
analysis attempts involving various techniques.  Features of the data
discovered in these attempts required changes in preprocessing and
strategy. These changes are described below as needed.

Each data set was analyzed by two or three methods for comparison
purposes. The simplest method is a conventional analysis based on
coincidence counting. The results of this method are susceptible to
the coincidence loophole and require strong assumptions on the source
and its statistics.  The second method involves our distance-based
Bell-function analysis applied to trials consisting of all the data
acquired while the settings were held fixed.  The third computes
``certificates'' of violation (given as $\log_{2}(p)$-values) using
the prediction-based-ratio (PBR)
protocol~\cite{zhang_y:qc2011a,Zhang2013} with truncated versions of
the distance-based Bell functions.  We discuss the analysis of the
three experiments in reverse order, which is also the order in which
the data sets were received and analyzed.

\subsection{Continuously Emitting Quantum Source}

The data set for this experiment consists of $3953$ trials with randomly
chosen settings. The measurement outcomes consist of a sequence of
timetags for each party, where each timetag records a detection
event. The average numbers of recorded detections per trial are approximately
$1400$ on setting $1$ and $4900$ on setting $2$ for both parties.
Each trial's results are stored in one file. The files for eight
trials were corrupted and therefore discarded, leaving $3945$ trials.
The timetag sequences were preprocessed in two steps.  The
preprocessing parameters were determined at an early stage of analysis
with a set consisting of $97\times 4$ randomly chosen trials with $97$
at each settings choice. (The final analysis was performed in the
order in which the experiment was performed with the initial trials
used for training--see below. The preprocessing parameters suggested
by the training set for the final analysis were the same up to
statistical fluctuations, so we did not change them for the final
analysis.)  The first preprocessing step compensated for transient
artifacts near the beginning and end of the timetag sequences. We
therefore used only timetags from the middle portion of the sequence
determined as follows: The sequence durations are approximately
$1\SI{s}$.  For each trial, we first determined the earliest recorded
time $t_{0}$ in both parties' sequences and set $t_{0}'$ to be the
second multiple of $10^{8}$ past $t_{0}$ (in the time units used for
the timetags, $156.25\SI{ps}$) Thus $t_{0}'=(\lfloor t_{0}/
10^8\rfloor+2)10^8$.  We then used only timetags with recorded times
$t$ satisfying $t_{0}'\leq t \leq t_{0}' + 6\times 10^{9}$.  We remark
that this preprocessing step is non-local, which is in general
undesirable. We are not aware of any way in which a local realistic
(LR) source could exploit this, though the possibility exists.  The
second preprocessing step corrected a systematic timing offset between
the recorded times for the two parties. The offset was applied to all
timetag sequences of party $A$ and involved shifting the timetags by
$-685$ time units (i.e., $107\SI{ns}$). For comparison, the time-jitter determined as the
typical distance between apparent coincidences is of the same order.

All analysis attempts used the preprocessing of the previous
paragraph.  We describe the final analysis first and then discuss how
we arrived at the final analysis. For the conventional analysis, we
used the first $197$ trials to determine the optimal coincidence
window. We then computed the number of coincidences for each trial.
The coincidences were determined as described in~\cite{KnillTBP}
rather than with the simple Alice-centered windows used in the main
text.  We then computed the estimated total violation as described in
App.~\ref{app:totviol} according to the Bell-inequality used for the
original analysis of the pulsed quantum source in
Ref.~\cite{Christensen2013}. The total violation according to this
analysis is $5.14(10)\times 10^4$, corresponding to a nominal
signal-to-noise ratio (SNR) of $59.7$. The latter is the ratio of the
total violation to the estimated uncertainty; see
App.~\ref{app:totviol}.

The distance-based analysis was performed adaptively. An adaptive
procedure was required because the parameters of the distance-based
Bell function used are sensitive to the apparent drifts in count-rates
over time.  Starting at the $201$st trial and then every $400$
trials, we re-optimized the Bell function parameters on the previous
$800$ trials. (Before the $801$st trial we used all the trials
already processed, including the first $200$.) The Bell function was
then computed for each of the next $400$ trials, before re-optimizing
the parameters. We then estimated the total violation as described in
App.~\ref{app:totviol}. The total violation according to the
distance-based analysis is $2.52(10)\times 10^{4}$, corresponding to a
nominal SNR of $24.5$.

The Bell-function values from the distance-based analysis were then used in an
adaptive version of the PBR analysis. This required adaptively
computing the parameters for Bell-function truncations and the
mixtures used in constructing the test factor according to the
protocol in Ref.~\cite{KnillTBP}.  This proceeded similarly to the
distance-based analysis, except that the parameters were updated every $10$
trials and optimized on the last $400$ (or less) trials. We chose the
more frequent update because there is little computational cost in
doing so and the truncation and mixture parameters are sensitive to
small drifts in the conditional means of the Bell function. The
$\log_{2}(p)$-value bound obtained is $269$, equivalent to a one-sided Gaussian
SNR of $19$.

The final analysis was performed after two previous analyses. The
first analysis involved partitioning the trials into randomly chosen
sets of four matched trials, one at each of the settings choices. The
version of the distance-based analysis available at the time yielded a
significantly smaller total violation than the final analysis.  The
PBR analysis at the time was faulty, but suggested a significantly
higher $\log_{2}(p)$-value bound than revealed by the final
analysis. Because the randomization strategy used in this analysis is
not acceptable for certification purposes, a second analysis was
performed after the analysis procedures were updated.  During this
analysis, we discovered that the count-rate variations in time
significantly affect the $\log_{2}(p)$-value bounds, requiring that
the analysis be performed adaptively. A choice for adaptation
parameters was made after investigating the timescale of the
variations.  The estimated total violation found was the same within
error bars as the one for the final analysis. The third and final
analysis was required because we discovered an error in our original
method for Bell function truncation in the PBR analysis resulting in
an overly optimistic $\log_{2}(p)$-value. The adaptation parameters
were chosen for the final analysis based on our experience in the
second round of analysis. Because of this history, a moderate bias in
the estimated total violation and in the $\log_{2}(p)$-value bounds is
expected.

\subsection{Pulsed Quantum Source}

The data set for this experiment consists of $4450$ trials.  The settings
for each trial were chosen randomly. Each party's measurement outcome
consists of a sequence of time-tagged detections. The source was
pulsed with the pulses synchronized with a clock whose ``ticks'' were
also recorded for each trial. There are approximately $12500$ pulses
per trial before preprocessing.  The original analysis of the data
reported in Ref.~\cite{Christensen2013} analyzed each pulse as a
trial. To consider the results of this analysis as evidence against LR
requires an assumption such as that each pulse is independent and
identical. Defining trials so that they contain all the
measurements that occur while the settings are fixed avoids making
this assumption.

The sequences of timetags from each trial were preprocessed as
follows: We first corrected for the time-offset of $A$ as we did for
the data from the continuously emitting source. We then removed
detections outside narrow windows containing each pulse.  The windows
were determined relative to the recorded clock ticks and have a width
of $16000$ time units.  The pulses are separated by about $256000$
time units. We dropped the first $200$ pulses and saved the $12200$
subsequent pulses, dropping the rest. To correct for intermittent
interference causing excess detections, we ``blanked out'' (removed
detections in) pulses where there was an excess number (three or more)
of detections outside pulse windows in the period spanning three
pulses before and after. Note that the preprocessing is local in the
sense that the parties can in principle perform it without
communicating, given that they have synchronized clocks.

As in the case of the continuously emitting source, there were several
analysis attempts. In the first attempt, the order of the trials
was randomized and we only confirmed that the violation based on
distance-based analysis was consistent with the results reported in
Ref.~\cite{Christensen2013}. The PBR analysis was not performed at this
time. Later analyses were performed in parallel with the analysis of
the continuously emitting source, with the final analysis correcting
the same problem with our original implementation of Bell-function
truncation.

For the final analysis, we did not perform a version of the
conventional coincidence analysis as the pulsed nature of the source
made this unnecessary. Applying the distance-based analysis using the
distance-based Bell functions of Ref.~\cite{KnillTBP} failed to show a
violation; we attribute this to the presence of an excess of
multiple detections during pulses and the sensitivity of the analysis
to detection-rate changes. (We attribute most multiple detections to
local detection artifacts, such as detector after-pulsing, rather than
photons. These local effects confuse the distance-based analysis by
adding non-violating LR signals.)  We therefore used a simpler
Bell-function with no parameters. This Bell-function is obtained by
adding the Bell-function derived from the Bell inequality used in
Ref.~\cite{Christensen2013} over the detections for each pulse.  For
this purpose, multiple detections in a pulse are counted as one.  This
is an instance of a general strategy for pulsed sources where the
settings are not changed for every pulse.  Consider a Bell function
$B$ for the detections from one pulse satisfying the Bell inequality
$\lang B(T,S)\rang_{\LR} \leq 0$, where $T$ is the detection pattern
and $S$ the measurement settings.  If we have a sequence of pulses at
fixed measurement settings $S$ with detection patterns $T_{i}$, we can
define a Bell function $B'(\mathbf{T},S)=\sum_{i}B(T_{i},S)$, where
$\mathbf{T}$ is the sequence of detection patterns $T_{i}$. The Bell
inequality $\lang B'(\mathbf{T},S)\rang_{\LR}\leq 0$ is also satisfied
by $B'$. To avoid assuming that trials are independent and identically
distributed (i.i.d.), one can analyze the violation of $B'$ instead of
$B$. This change of view enables the PBR analysis: As noted in
Ref.~\cite{KnillTBP}, all Bell functions for two parties and two
settings can be derived from a set of distance-like functions
satisfying an iterated triangle inequality. This makes it possible to
apply the PBR analysis as we have done here.  Although the parameters
of the Bell-function require no training, the total violation was
computed using the procedure of App.~\ref{app:totviol} with an initial
set of $200$ trials set aside for initializing the predictions. In
each step, the predictions in the procedure were updated using the
previous $200$ trials to account for experimental drift.  The
distance-based analysis found a total violation of $1.41(18)\times
10^{3}$, corresponding to a nominal SNR of $7.8$.

For the PBR analysis, we computed the necessary truncation and mixture
parameters adaptively based on the Bell function values obtained in
the Bell function analysis. The first $400$ trials were reserved for
training.  Starting with the $401$st trial, we updated the parameters
every $200$ trials based on the previous $1600$ trials (or less,
initially). We found a $\log_{2}(p)$-value bound of $33$, equivalent
to a one-sided Gaussian SNR of $6.3$.

\subsection{Classical Source}

The experiment on the classical source consisted of $9$ groups of four
trials at each of the four settings choices. The settings were not
chosen randomly. Thus the interpretation of an apparent violation
requires i.i.d. assumptions. Of course, the source was designed to be
LR, so no real violation can be observed.  The data from this
experiment was analyzed just once, after the distance-based analysis
matured.  Each trial has approximately $97000$ detections for each
party independent of the settings. The first group was set aside for
training. A first step in all our analyses was to determine systematic
timing offsets and an estimate of the time-jitter, both were done by checking
timetag differences on apparent coincidences.  For this source, the
timetag differences immediately revealed that there was an
``unexpected'' pattern in the detections. That is, since there was no
attempt at hiding that the source was exploiting the coincidence-time
loophole, the resulting characteristic detection delays are
obvious. (Ref.~\cite{KnillTBP} demonstrates a simulated source that
can successfully hide these detection patterns.)

For the purpose of demonstrating that a standard coincidence analysis (windows determined by Alice's detections)
is deceived by this source, we optimized the coincidence window as
usual on the training set and applied the coincidence analysis to the
rest. The total violation found was $6.6488(24)\times 10^5$ for a large
nominal SNR of $2781$. We optimized the Bell function for the distance-based
analysis but were unable to detect a violation. In fact, the estimated
total Bell function was significantly negative.  We cannot exclude the
possibility that a better choice of parameters for the timetag Bell
function exists, though we know on theoretical grounds that a
violation should not be observable according to the distance-based
analysis. Given the absence of violation, the PBR analysis is
guaranteed to use trivial test factors giving a $\log_2(p)$-value
bound of $0$.

\section{Conservative Estimates of Bell-Violation}
\label{app:totviol}

For each experiment, the timetag Bell function $B$ used has
expectations that are related to the violation of a CH-type inequality
by multiplying the latter by the expected number of photon pairs.  In
the limit of low time-jitter compared to the mean photon-pair
inter-arrival time, the expectations according to $B$ and that
expected from the CH-type inequality converge. It is therefore
worthwhile to estimate the expected value of the Bell function for
comparison purposes. In principle, for each trial, the expectation of
$B$ is an experimental observable that, if greater than zero,
witnesses violation of the Bell inequality associated with $B$.  If
the trials are i.i.d., the expectation can be estimated empirically
using conventional methods.  For tests of LR, this is usually done by
estimating the settings-conditional expectations of $B$, which are
then added.  The uncertainty is obtained accordingly. When the trials
are not necessarily independent or identical, there may be no single
expectation of $B$ to estimate, so the conventional method cannot be
used.  Here we give an alternative that produces meaningful results in
the general case. It statistically agrees with the conventional method
when the trials are i.i.d.: While the estimated uncertainties obtained
are slightly larger on average, they differ by an amount that is
comparable to the expected statistical fluctuations in the estimate.

We emphasize that the purpose of these methods is to obtain an
estimate of a physical quantity and the associated uncertainty. They
do not yield certificates against LR (see~\cite{Zhang2013} for a
discussion). While we obtain uncertainties that are appropriate for
dependent trials whose expectations change in time under normal
experimental conditions, a sufficiently determined adversary can still
ensure that our uncertainties are overly optimistic.

Here is the procedure for our method. A mathematical discussion
follows the procedure. 
\begin{itemize}
\item[1.] 
  \begin{itemize}
  \item[a.] Initialize the running value of the estimated total Bell
    violation by setting $\hat b_{[0]} = 0$.
  \item[b.] Initialize the running value of the estimated variance
    $\hat u_{[0]} = 0$
  \end{itemize}
\item[2.] For each trial result $d_{i}$, $1\leq i\leq N$ in order, do
  the following
  \begin{itemize}
  \item[a.] Before considering $d_{i}$:
    \begin{itemize}
    \item[1.] Predict the settings-conditional expected
      Bell-function values $\langle B(D_{i})|S_{i}=s\rangle$ at the
      next trial as $b_{\mathrm{pred},i}(s)$. This prediction can be
      based on any information available before the $i$th trial
      occurred, including calibrations, theory and previous trial
      results.  Here, $S_{i}=(S^{A}_{i},S^{B}_{i})$ are the joint
      settings random variables and $D_{i}$ are the random variables
      whose outcome values are the $d_{i}$.
    \item[2.]  Determine the predicted average Bell function
      violation $\bar b_{\mathrm{pred},i}=\sum_{s}
      b_{\mathrm{pred},i}(s) p_{s}$, where $p_{s}$ is the
      probability of settings choice $s$. Note that 
      $\langle b_{\mathrm{pred},i}(S)\rangle=\bar
      b_{\mathrm{pred},i}$ is known exactly before the $i$th trial.
    \end{itemize}
  \item[b.] Now consider $d_{i}$:
    \begin{itemize}
    \item[1.] Compute $\hat b_{[i]} = \hat
      b_{[i-1]}+B(d_{i})-(b_{\mathrm{pred},i}(s_{i}) - \bar
      b_{\mathrm{pred},i})$.
    \item[2.] Compute $\hat u_{[i]} = \hat u_{[i-1]}+ (B(d_{i})-b_{\mathrm{pred},i}(s_{i}))^{2}$.
    \end{itemize}
  \end{itemize}
\item[3.] Report the estimated total Bell violation as $\hat
  b_{[N]}$ with an approximate $68\,\%$ confidence
  interval of $\hat b_{[N]}\pm\sqrt{\hat u_{[N]}}$.
\end{itemize}

The simplest method for predicting the settings-conditional
Bell-function expectations in step 2.a.1 of the procedure is to
compute the sample means conditional on settings from the first
$i-1$ trials and the training trials. This works well for stable
experiments. For the data analyses in this paper, we used a segment
of recent trials (including the training trials) instead.  We
formulated the procedure for a fixed Bell function, but the
procedure also works if the Bell functions are chosen
adaptively before each trial.

Consider a sequence of trials with each trial's result given by
$d_{i}$.  We now adopt the usual conventions for random variables and
their outcome values, where random variables are capitalized. Thus
$D_{i}$ is the random variable for the result from the $i$th trial,
and $d_{i}$ is its outcome value in a particular run of the
experiment.  The results consist of the measurement outcomes and
settings. We let $T^{X}_{i}$ and $S^{X}_{i}$ be the respective random
variables for the measurement outcome and setting of party $X$ in the
$i$th trial.  We let $D$ denote the sequence of random variables
$D_{i}$.  The random variables $D_{i}$ are not necessarily
independent, but the distributions of the settings $S^{X}_{i}$ are
jointly uniform and therefore independent of each other.  We let
$H_{i-1}$ be a random variable that captures the history of events
preceding trial $i$, including events not captured by $D$ but that are
relevant to the experiment. In particular, $H_{i-1}$ determines
$D_{j}$ for $j<i$ and may include additional experimentally relevant
information. The goal is to obtain an empirical estimate of the
quantity
\begin{equation}
  \bar b_{[N]}(h) = \sum_{i=1}^{N}\langle B(D_{i})|H_{i-1}=h_{i-1}\rangle,
\end{equation}
and a confidence interval for this estimate.  Here, $h_{i-1}$ is the
actual value of the history random variable.  Throughout, we assume
that the relevant real-valued random variables have finite second
moments.  We interpret $\bar b_{[N]}(h)$ as the total Bell inequality
violation actually present in the experiment, which we estimate with
$\hat b_{[i]}$.  We do not assume that the outcome value $h_{i-1}$ is
known, just that it is well-defined for a given run of the
experiment. Define $\bar b_{i}(h) = \bar b_{i}(h_{i-1}) = \langle
B(D_{i})|H_{i-1}=h_{i-1}\rangle$. This is the expected value of the
Bell function for the upcoming $i$th trial, just before the trial is
performed.  We use the following conventions to refer to functions of
random variables and the random variables defined by these functions:
Except for the Bell function $B$, we use lower case annotated symbols
for the functions. Applying a function to a random variable as in the
expression $\bar b_{i}(H)$ defines a new random variable. To simplify
the notation, we also refer to this random variable by its upper case
variant, so that $\bar B_{i} = \bar b_{i}(H)$.  (Here $H$ refers to
the full history.)  The outcome values of this random variable are
then denoted by $\bar b_{i} = \bar b_{i}(h)$.

For i.i.d. trials, $\bar b_{i}(h) = \langle
B(D_{i})\rangle$ and is independent of $i$ and $h$.  The sum $\bar
b_{[N]}(h)=\sum_{i=1}^{N}\bar b_{i}(h)$ can then be interpreted as the
conventional total Bell inequality violation of the experiment, if it
is positive.  For the empirical estimate of $\bar b_{[N]}(h)$ we could
compute $\sum_{i=1}^{N}B(d_{i})$, but instead we use the less-noisy
estimate $\hat b_{[N]}$ from the procedure.  The estimate is less
noisy because we subtracted from $b(d_{i})$ the quantity
$(b_{\mathrm{pred},i}(s_{i}) - \bar b_{\mathrm{pred},i})$, whose mean
is guaranteed to be zero but is expected to be positively correlated
with the original estimate.

The first task is to show that $\hat b_{[N]}$ is an unbiased estimator
of $\bar b_{[N]}(h)$ (that is $\langle\hat B_{[N]}\rangle = \langle
\bar b_{[N]}(H)\rangle$). We use $E(A|B)$ to denote the
conditional expectations of $A$ with respect to $B$ interpreted as a
function of the random variable $B$. The notation
$\langle\ldots\rangle$ is reserved for unconditional expectations and
expectations conditional on specific outcome values.  Since
$E(\ldots|\ldots)$ denotes random variables, they may occur inside
$\langle\ldots\rangle$.  By expanding the definition, we have
\begin{equation}
  \left\langle \hat B_{[N]} \right\rangle = \left\langle \sum_{i=i}^{N}B(D_{i})-(B_{\mathrm{pred},i}-\bar B_{\mathrm{pred},i}) \right\rangle .
\end{equation}
The expectation of $B_{\mathrm{pred},i} - \bar B_{\mathrm{pred},i}$ is
$0$ by design, so
\begin{eqnarray}
  \left\langle \hat B_{[N]} \right\rangle &=& \sum_{i=i}^{N} \left\langle B(D_{i}) \right\rangle \\
  &=& \sum_{i=i}^{N} \left\langle E(B(D_{i})|H_{i-1}) \right\rangle \\
  &=& \sum_{i=i}^{N} \left\langle \bar b_{i}(H) \right\rangle \\
  &=& \left\langle \bar b_{[N]}(H) \right\rangle,
\end{eqnarray}
where the identity $\left\langle B(D_{i}) \right\rangle = \left\langle
  E(B(D_{i})|H_{i-1}) \right\rangle$ follows from the rules
for iterated conditional expectations. (This is a special case
sometimes referred to as the ``law of total expectations''.)

The second task is to determine an approximate $68\,\%$ confidence
interval for $\bar b_{[N]}(h)$ around $\hat b_{[N]}(d)$.  Note that
the confidence interval is itself a random variable with respect to
$H$ that should reflect what actually happened during the experiment
as indicated in the definition of $\bar b_{[N]}(h)$. Formally, we seek
a bound $\underline \delta$ for a (conservative) confidence interval
for $\bar b_{[N]}(h)-\hat b_{[N]}(d)$ that satisfies a coverage
condition, namely that \emph{before} the experiment, the probability
that $-\underline\Delta\leq \bar b_{[N]}(H)-\hat b_{[N]}(D)\leq
\underline\Delta$ is at least $68\,\%$.  Here $\underline\Delta$ is
the random variable with outcome values $\underline\delta$. (We could
consider $-\underline\delta$ as the lower endpoint of a one-sided
confidence set with no upper bound, in which case we require that
\emph{before} the experiment, the probability that $-\underline\Delta
\leq \bar b_{[N]}(H)-\hat b_{[N]}(D)$ is at least $84\,\%$.)  Because
the trials may not be i.i.d., the standard estimates of variance
cannot be applied to determine $\underline \delta$.  Our method yields
an estimate of an error bound given relatively mild assumptions and
sufficiently large $N$.

Let $\hat b_{i}
= B(d_{i}) - b_{\mathrm{pred},i}(s_{i})+\bar b_{\mathrm{pred},i}$ be
the increment $\hat b_{[i]}-\hat b_{[i-1]}$ of the estimated Bell
violation from the $i$th trial.  By design of $b_{\mathrm{pred},i}$, we have
$E(\hat B_{i}|H_{i-1}) = E(B(D_{i})|H_{i-1}) = \bar b_{i}(H)$.

We investigate the statistics of the estimation error $\Delta_{[N]} =
\hat b_{[N]}(D)-\bar b_{[N]}(H) = \sum_{i=1}^{N}\Delta_{i}$, with
\begin{equation}
  \Delta_i=\hat B_{i}-\bar B_{i}.
\end{equation}
Note that $\langle \Delta_{i} \rangle = 0$ and $\langle \Delta_{[N]}=0
\rangle$, so the variance of $\Delta_{[N]}$ is
\begin{equation}
  \mathrm{Var}(\Delta_{[N]}) = \left\langle \left(\sum_{i=i}^{N} \Delta_{i}\right)^{2}\right\rangle.
\end{equation}
Since $E(\Delta_i|H_{i-1})=0$, the $\Delta_i$ are martingale
increments adapted to the $H_i$. (For the relevant theory of
martingales, see Ref.~\cite{ph_marts}.)  Martingale increments at
different times are uncorrelated. That is, for $i>j$, $\langle
\Delta_i\Delta_j\rangle = \langle E(\Delta_i\Delta_j|H_j)\rangle =
\langle E(\Delta_i|H_j)\Delta_j\rangle = 0$. A consequence is that the
variance of the estimation error satisfies $\mathrm{Var}(\Delta_{[N]})
= \left\langle\sum_{i=1}^{N}\Delta_i^{2}\right\rangle$. In detail,
\begin{eqnarray}
  \left\langle \left( \sum_{i=1}^{N}\Delta_{i} \right)^{2} \right\rangle 
  &=& \left\langle \sum_{i>j}2\Delta_{i}\Delta_{j}+\sum_{i=1}^{N}\Delta_{i}^{2}\right\rangle \\
  &=& \sum_{i>j} 2\langle\Delta_{i}\Delta_{j} \rangle + \left\langle \sum_{i=1}^{N}\Delta_{i}^{2} \right\rangle \\
  &=&  \left\langle \sum_{i=1}^{N}\Delta_{i}^{2} \right\rangle.
\end{eqnarray}

Since we do not know $\bar b_{i}(h)$, we cannot directly compute
$\delta_{i}^{2}$ as our estimate of $\langle
\Delta_i^2|H_{i-1}=h_{i-1}\rangle$. But we can use the prediction $\bar
b_{\mathrm{pred},i}$ of $\bar B_{i}(h)$ made before the $i$th trial.
Recall that the variance of a random variable $R$ is the minimum
expectation of $(R-m)^{2}$, where the minimum is achieved by
$m=\langle R\rangle$. In conditional form, this implies
$E(\Delta_{i}^{2}|H_{i-1})\leq E((\Delta_{i}-M)^{2}|H_{i-1})$ for
any $M$ that is a function of $H_{i-1}$, because
$\Delta_{i}$ is zero-mean conditional on $H_{i-1}$.  We set $M=
\bar B_{\mathrm{pred},i}-\bar B_{i}$ and define
\begin{equation}
\hat\delta_{i} = \delta_{i} - m = B(d_i)-b_{\mathrm{pred},i}(s_{i}),
\end{equation}
which we can compute from the available information.
The variance inequality now implies
$E(\hat\Delta_{i}^{2}|H_{i-1})\geq
E(\Delta_{i}^{2}|H_{i-1})$, so $\hat u_{[N]} =
\sum_{i=1}^{N}\hat\delta_{i}^{2}$ can serve as a biased-high estimate
of the desired variance. Formally,
\begin{eqnarray}
  \langle\hat U_{[N]}\rangle 
  &=& \sum_{i=1}^{N}\langle \hat\Delta_{i}^{2}\rangle \\
  &=& \sum_{i=1}^{N}\langle E(\hat\Delta_{i}^{2}|H_{i-1})\rangle\\
  &\geq& \sum_{i=1}^{N}\langle E(\Delta_{i}^{2}|H_{i-1})\rangle \\
  &=& \sum_{i=1}^{N}\langle\Delta_{i}^{2}\rangle\\
  &=& \mathrm{Var}(\Delta_{[N]}),
\end{eqnarray}
where $\hat U_{[N]}$ is the random variable corresponding to $\hat
u_{[N]}$.  

To justify $\sqrt{\hat u_{[N]}}$ as an estimated uncertainty requires
additional assumptions on the random variables. For Chebyshev-type
inequalities involving variance and a variety of exponential bounds
on tail probabilities, boundedness of $B$ suffices (and is
typically stronger than necessary). But one would like to use
appropriate central-limit theorems in the same way as for
i.i.d. trials.  The conditions under which such central-limit
theorems hold are surprisingly broad, but not unrestricted.
Ref.~\cite{ph_marts} has a variety of relevant versions that can be
applied in non-adversarial situations where the square errors
$\Delta_{i}^{2}$ are asymptotically well-behaved. We therefore
suggest that in typical physics experiments with sufficiently many
trials without excessive stability problems, the approximate
$68\,\%$ confidence interval of the total Bell
violation can be given as $\hat b_{[N]}\pm\sqrt{\hat u_{[N]}}$.
We expect this interval to be conservative under most conditions
even though the trials need not be i.i.d.

\bibliographystyle{unsrt}
\bibliography{BibBTCT}

\end{document}